\documentclass[journal]{IEEEtran}
\IEEEoverridecommandlockouts

\usepackage{amsmath}
\usepackage{graphicx}
\usepackage{cite}
\usepackage{amsfonts}
\usepackage{algorithm}
\usepackage{algorithmic}

\begin{document}

\title{Enhanced Secure Transmission of Medical Images through OFDM using Hyperchaotic Systems}

\author{
    \IEEEauthorblockN{Nada Bouchekout\IEEEauthorrefmark{1}, Abdelkrim Boukabou\IEEEauthorrefmark{1}, Morad Grimes\IEEEauthorrefmark{2}}\\
    \IEEEauthorblockA{\IEEEauthorrefmark{1}Laboratory of Renewable Energy, Department of Electronics, University of Jijel,\\ BP 98 Ouled Aissa 18000, Jijel, Algeria} \\
    \IEEEauthorblockA{\IEEEauthorrefmark{2,1}Laboratory of Non-Destructive Testing, Department of Electronics, University of Jijel, \\
    BP 98 Ouled Aissa 18000, Jijel, Algeria }\\
    Email: nada.bouchekout@univ-jijel.dz, aboukabou@univ-jijel.dz, moradgrimes@univ-jijel.dz}
\markboth{Fourth International Conference On Technological Advances in Electrical Engineering (ICTAEE'23), May 23-34 2023.}{}

\maketitle

\begin{abstract}
Orthogonal Frequency Division Multiplexing (OFDM) is a popular modulation technique for transmitting digital data over wireless radio channels, including medical images due to its high transmission capacity, low interference, bandwidth efficiency, and scalability. However, the security of medical images is a major concern, and combining OFDM with encryption techniques such as chaos-based image encryption can enhance security measures. This study proposes a secure medical image transmission system that combines OFDM, 6D hyperchaotic system, and Fibonacci Q-matrix and analyzes its impact on image transmission quality using simulation results obtained through MATLAB. The study examines the Q-PSK constellation diagram, fast Fourier transform (IFFT) signal, cyclic prefix (CP) techniques, NIST, signal noise ratio (SNR), and bit error rate (BER). The results provide insights into the effectiveness of OFDM in securely transmitting high-quality medical images.
\end{abstract}


\section{Introduction}
In recent years, data security has become increasingly important for data transmission. Medical images are commonly used in various processes \cite{eldin2014optimized}, and different methods and technologies such as data hiding \cite{lin2016novel}, steganography \cite{dragoi2015local}, and encryption \cite{hosny2021new} have been developed to protect digital images. Among these methods, image encryption is the most straightforward because it converts meaningful images into unrecognizable noise-like images, making them unintelligible to the human eye \cite{bouridah2017image}. Chaos theory is often used in cryptography due to its unique properties such as periodicity, sensitivity to initial conditions, and random-like behavior which meet the requirements of cryptography \cite{bouridah2017image}. As a result, a huge number of chaos-based encryption techniques have been proposed such as the logistic-sine map proposed by Chen and Hu \cite{chen2017adaptive} for medical image encryption, the memristive chaotic system implemented by Chai et al. \cite{chai2018image}, and the image encryption using a hyperchaotic system proposed by Hosney et al. \cite{hosny2021new}.  

In addition, the evolution of technology has led to the development of OFDM as the suggested method for transmitting digital data over a wireless radio channel due to its high transmission capacity, bandwidth efficiency, reduced interference, system flexibility, and scalability \cite{naik2017efficient}. OFDM splits transmissions into sub-channels, enabling high bit rates and spectrum efficiency \cite{himeur2017robust}. In \cite{helmy2018chaotic}, the authors applied the OFDM system in combination with chaotic baker chart permutation to transmit encrypted images.  In \cite{naik2017efficient,truong2020performance}, the method of encrypting direct QAM symbols is discussed in terms of security.  In \cite{dharavathu2020secure},  the authors developed a method of encryption and scrambling on both sides for an OFDM system. The authors \cite{eldin2014optimized} have suggested an AWGN crypto-OFDM system for secure transmission. In \cite{ayad2022secure}, the transmission of secure images has been discussed regarding the logistic sine cosine algorithm through the OFDM communication system.

Motivated by the discussions above, in this paper, we present a study focusing on a secure medical image transmission system that uses the OFDM modulation method. This system incorporates a six-dimension (6-D) hyperchaotic system to combat the effects of multipath and enhance medical image transmission security. As a result, the proposed system can use OFDM modulation for the secure transmission of medical images providing a reliable and secure method for transmitting sensitive patient data. By employing strong encryption algorithms and channel coding techniques, medical professionals can ensure that patient data remains confidential and secure during transmission.

The structure of the paper is as follows: In the next section, we provide a brief explanation of OFDM modulation. In Section 3, we describe the proposed 6-D hyperchaotic system for the encryption of medical images. In Section 4, we present the results and discussion to demonstrate the effectiveness of the proposed method. Finally, we provide some conclusions in Section 5.

\section{Proposed System Model}
\subsection{OFDM system}
In this study, we investigate the secure transmission of medical images using an OFDM system as depicted in Fig.\ref{fg1}. First, the original medical image is encrypted with a 6-D hyperchaotic sequence, and the resulting data is converted into a bit stream. The bit stream is then mapped using QPSK to generate a sequence $[x = x_0, x_1, x_2, \dots, x_{N-1}]$ which is then passed through the IFFT modulator to produce a corresponding continuous-time signal. A cyclic prefix (CP) is added to each data block to mitigate inter-symbol interference (ISI), and the signal is converted to analog using a digital-to-analog converter \cite{himeur2017efficient}.

The transmitted signal $s(t)$ is given by:
\begin{equation}
s(t) = \frac{1}{N} \sum_{k=0}^{N-1} x(k) e^{j 2 \pi k t / T_s}, \quad 0 < t < T_s
\end{equation}

where $T_s$ is the active symbol interval, and $x(k)$ is the mapped version of the data sequence. The signal is then transmitted through an AWGN channel, which introduces Gaussian noise with zero mean and unity variance \cite{himeur2017efficient}.

\begin{figure}[]
    {\includegraphics[width=1\columnwidth]{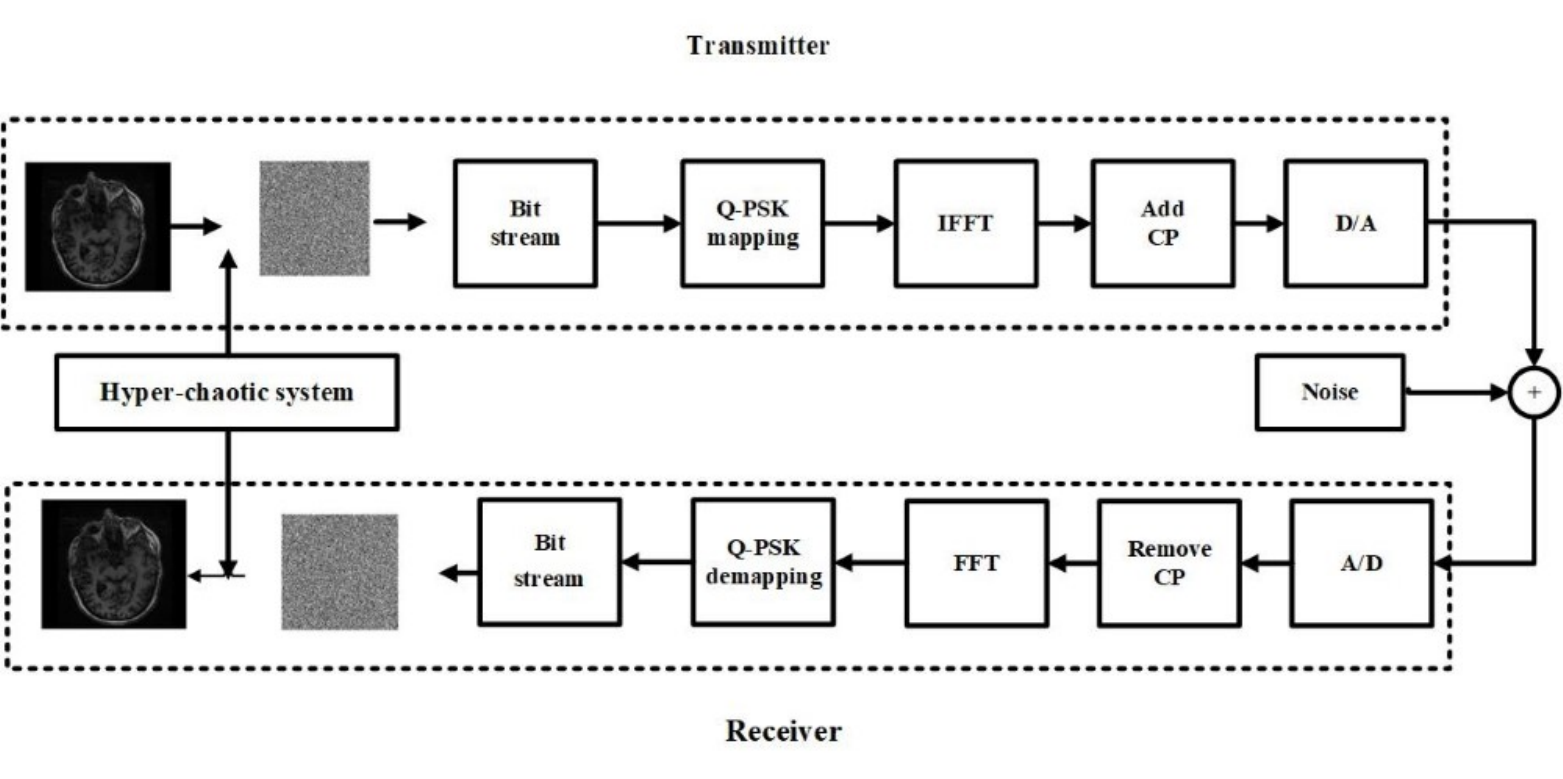}
  \label{co}}
    \caption{Block diagram of the OFDM baseband transmission model}
   \label{fg1}
\end{figure}

The received signal is processed by removing the CP and transforming it into the frequency domain using FFT. The data is then demapped and decrypted using a 6-D hyperchaotic system. Finally, the image decryption process is performed to reconstruct the original medical image.


\section{Hyperchaotic system}
The literature on chaos theory inspired researchers in this field of interest to conduct meaningful studies. One notable contribution is the paper by Lü and Chen which presents a 5-D hyperchaotic system. Another technique introduced by WANG et al. in \cite{wang2019new} is used to construct a 6-D hyperchaotic system by incorporating a nonlinear state controller into the existing system. The result is a 6-D hyperchaotic system, given by

\begin{equation}
\begin{aligned}
    \dot{x}_1 &= a(x_2 - x_1) + x_4 + x_5 - x_6, \\
    \dot{x}_2 &= c x_1 - x_2 - x_1 x_3, \\
    \dot{x}_3 &= -b x_3 + x_1 x_2, \\
    \dot{x}_4 &= d x_4 - x_2 x_3, \\
    \dot{x}_5 &= e x_6 + x_3 x_2, \\
    \dot{x}_6 &= r x_1,
\end{aligned}
\end{equation}
where $a$, $b$, $c$, $d$, $e$, and $r$ are the system parameters, and $x_1, x_2, x_3, x_4, x_5, x_6$ represent the state variables of the 6D \textit{hyperchaotic} system. The parameters are taken as $a = 10$, $b = 8/3$, $c = 28$, $d = -1$, $e = 8$, $r = 3$.

\subsection{Lyapunov exponents}
The Lyapunov exponent is a fundamental concept in chaos theory that measures the rate of divergence or convergence of trajectories in a nonlinear dynamic system. It indicates how much the system's behavior is affected by small changes in its initial conditions. The Lyapunov exponent spectrum is a useful method to assess a system's sensitivity to its starting point over time, and it is crucial in identifying the presence of chaotic attractors in the system \cite{wang2019new}.

Fig. 2 shows the Lyapunov exponent of the 6-D hyperchaotic system for which the initial conditions were set as (1,1,1,1,1,1).

\begin{figure}[]
    {\includegraphics[width=1\columnwidth]{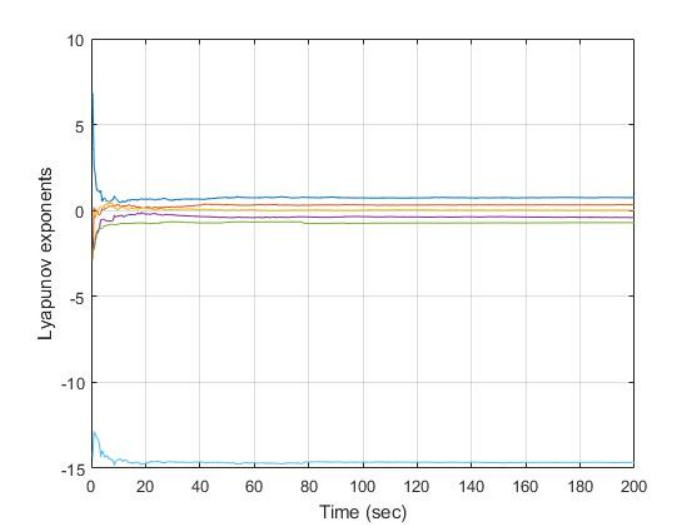}
  \label{co}}
    \caption{Lyapunov exponent of the system.}
   \label{fg1}
\end{figure}


\subsection{Bifurcation diagram}

\begin{figure}[]
    {\includegraphics[width=1\columnwidth]{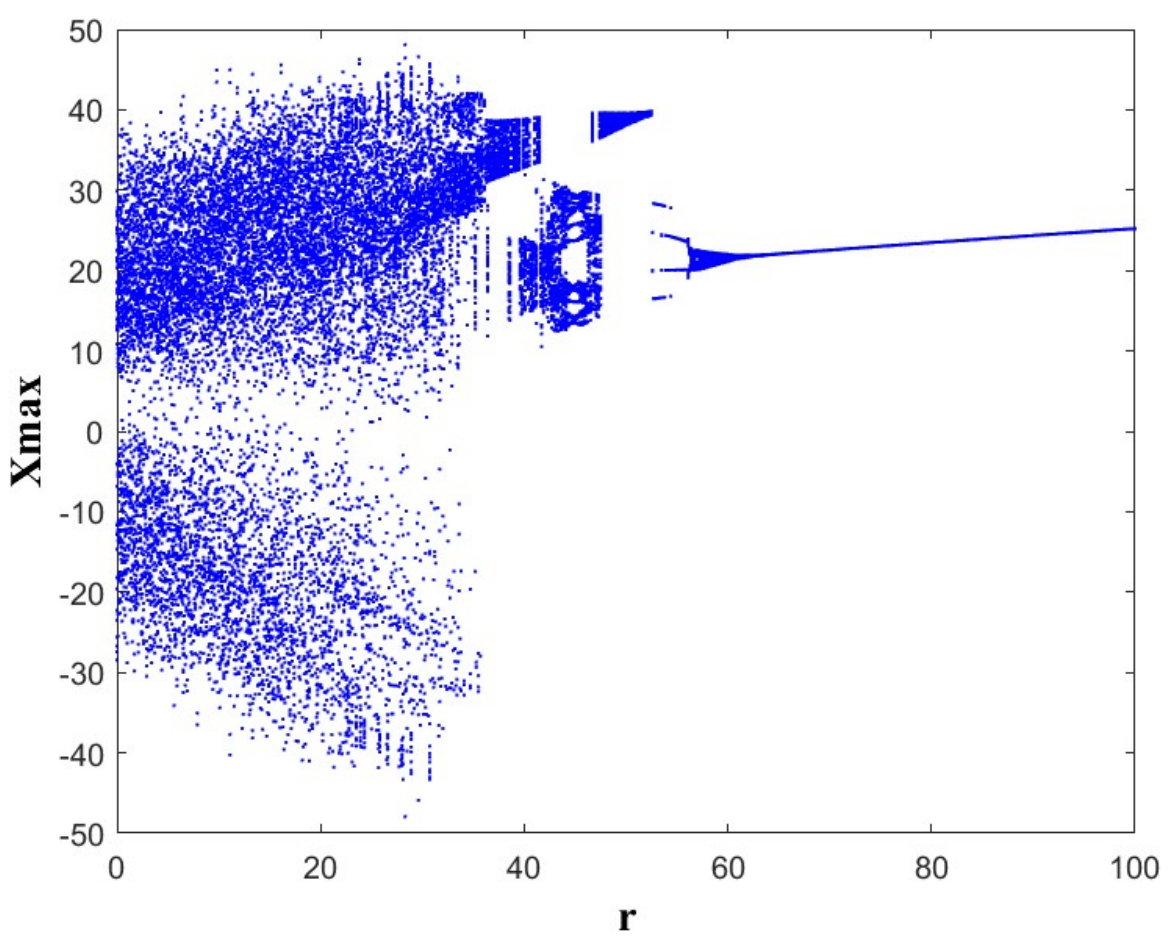}
 }
    \caption{Bifurcation diagram of the system.}
   \label{fg3}
\end{figure}

In mathematics, particularly in dynamical systems, a bifurcation diagram is an important tool for examining chaotic systems \cite{louzzani2021novel}. To investigate the dynamic behavior of the proposed 6-D hyperchaotic system, different values of the parameter r are tested, using an initial condition of ($x_1, x_2, x_3, x_4, x_5, x_6$)=(1,1,1,1,1,1)  with the other coefficients held constant. The bifurcation diagram is a crucial technique for revealing the behavior of chaotic systems, and the results are illustrated in Fig.\ref{fg3}. The diagram indicates that $X_{max}$ shifts from an unstable state to a stable state as the linear coefficient $r$ increases, indicating significant changes in the system.

\subsection{Image Encryption}

The input image was encrypted using an algorithm that utilized a 6-D hyperchaotic system and a Fibonacci Q-matrix. The use of the 6-D hyperchaotic system is advantageous due to its highly dynamic and complex behavior, as well as its two positive Lyapunov exponents which result in improved encryption performance and increased security. The Fibonacci Q-matrix is a simple and fast method that effectively diffuses the scrambled image \cite{hosny2021new}. Accordingly, the medical image encryption algorithm is performed using Algorithm 1.

\begin{algorithm}
\caption{A summary of the medical image encryption algorithm.}
\begin{algorithmic}
    \STATE \textbf{Step 1.} Set $i = 1$
    
    \STATE \textbf{Step 2.} Convert the medical image array to a vector $P$.
    
    \STATE \textbf{Step 3.} Determine the initial key for the \textit{hyperchaotic} system in the following manner:
    \begin{equation}
    x_1 = \frac{\sum_{i=1}^{M \times N} P(i) + (M^3 \times N)}{2^{8 \times (M^2 + N)}}
    \end{equation}
    
    \begin{equation}
    x_i = \text{mod}(x_{i-1} \times 10^6, 1), \quad i = 2,3,\dots,6
    \end{equation}

    With the initial conditions as $x_1, x_2, \dots, x_6$. Next, iterate the \textit{hyperchaotic} system in equation (2) for $\frac{N_0 + M \times N}{3}$ times, and then discard the first $N_0$ values to obtain the desired result.
    
    \STATE \textbf{Step 4.} Create a new sequence $L$ of size $M \times N$ by selecting three sequences ($x_1$, $x_3$, and $x_5$) from the system in equation (2) and sort $L$ in ascending order and record their positions in vector $S$.
    \STATE \textbf{Step 5.} Use the image vector $P$ to generate a newly shuffled sequence $R$ according to the following rule:
    \begin{equation}
    R_i = P\left(S_i\right), \quad i = 1: M \times N
    \end{equation}
    
    \STATE \textbf{Step 6.} Transform the sequence $R$ into the matrix $R_0$ and split it into sub-blocks of size $2^8 \times 2^8$. Compute the Chipped image $C$ by multiplying each $2^8 \times 2^8$ sub-block in $R_0$ with the Fibonacci Q matrix $(Q^{20})$.
    
    \STATE \textbf{Step 7.} Let $I = C$, then $i = i + 1$.
    
    \STATE \textbf{Step 8.} Replicate Step 2 to Step 6 for $i \leq 4$.
    
    \STATE \textbf{Step 9.} Chipper the medical image $C$.
    
    \STATE \textbf{Step 10.} Modulate chipped medical image using OFDM introduced in section 2.
\end{algorithmic}
\end{algorithm}

The resulting chipped image C is the encrypted medical image. To decrypt the image, the reverse steps can be performed in the same order using the same secret key.

\section{Results and Discussion}
\subsection{OFDM simulation}
\begin{figure}[]
{\includegraphics[width=1\columnwidth]{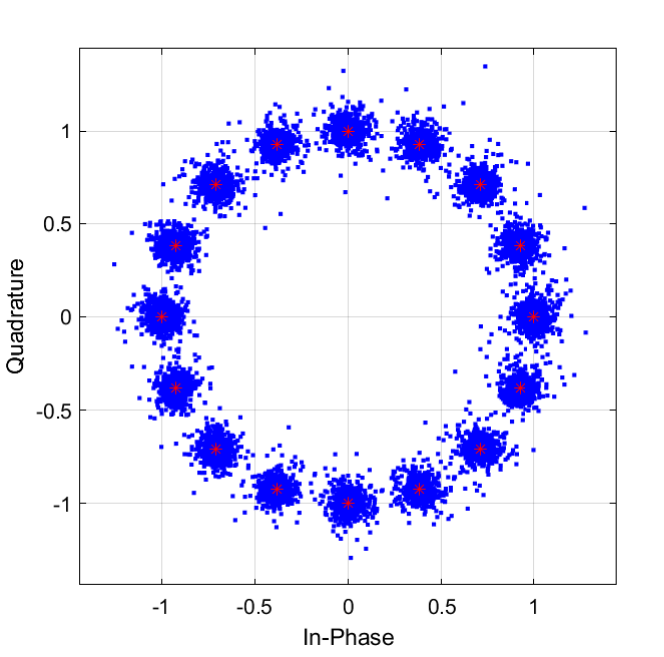}}
    \caption{Constellation diagram 16-PSK.}
   \label{fg4}
\end{figure}

This section presents the results of MATLAB simulations that investigate the effect of Orthogonal Frequency Division Multiplexing (OFDM) on secure medical image transmission. We provide an overview of the simulation parameters in Table\ref{tab:path2}, and show the 16-PSK modulation constellation diagram used in the OFDM technique in Fig. \ref{fg4}. The use of 16-PSK modulation provides a higher data rate which is important in medical image transmission.

\begin{table}[]
\caption{Parameters of the simulation proposed system}
    \label{tab:path2}
\fontsize{1\baselineskip}{2\baselineskip}\selectfont
\begin{center}
\begin{tabular}{c c }
\hline
\textbf{Parameters}           & \textbf{Simulation}  \\
\hline
\textbf{Image size}           & 256*256              \\ 
\textbf{FFT length}           & 1024                 \\ 
\textbf{Modulation}           & OFDM                 \\ 
\textbf{Mapping}              & Q-PSK                \\ 
\textbf{Cyclic prefix length} & 256                  \\ 
\textbf{Noise}                & AWGN \\ \hline
\end{tabular}
\end{center}
\end{table}

\begin{figure}[]
    {\includegraphics[width=1\columnwidth]{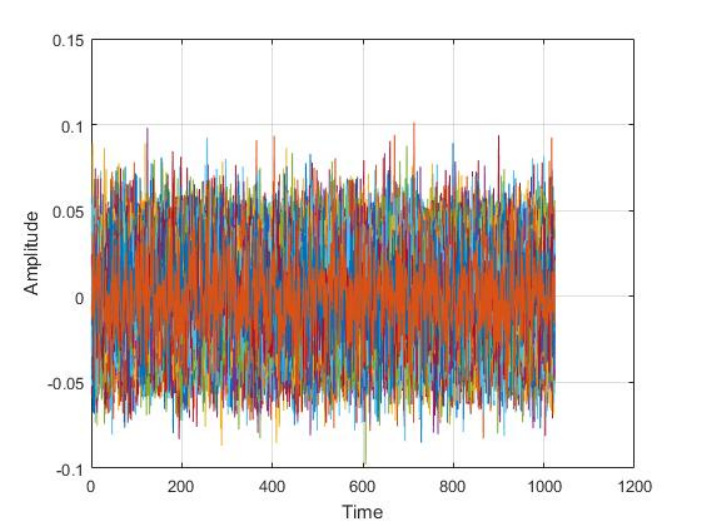}}
    \caption{Visualization of OFDM Signal After  Inverse Fast Fourier Transform ( IFFT ) Processing.}
   \label{fg5}
\end{figure}

\begin{figure}[]
    {\includegraphics[width=1\columnwidth]{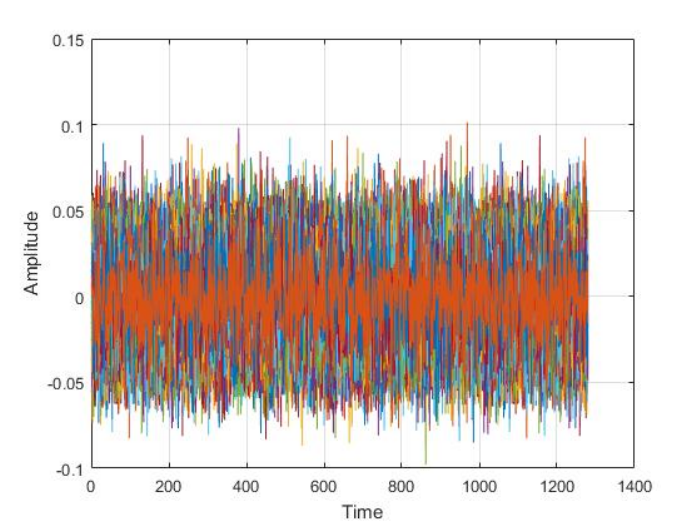}}
    \caption{Improving OFDM signal transmission with cyclic prefix (CP).}
   \label{fg6}
\end{figure}

Our proposed scheme improves the OFDM system performance by using IFFT and CP techniques. Fig.\ref{fg5} shows the IFFT operation that converts the frequency-domain signal into the time-domain signal, allowing for the simultaneous transmission of multiple subcarriers. Fig.\ref{fg6} illustrates the use of CP which is inserted at the beginning of each OFDM symbol to mitigate the effects of channel delay spread. Our study aims to enhance transmission efficiency and ensure the confidentiality of medical data, and our results demonstrate that the proposed techniques can improve the performance of the OFDM system for secure medical image transmission. 

\begin{figure}[H]
{\includegraphics[width=1\columnwidth]{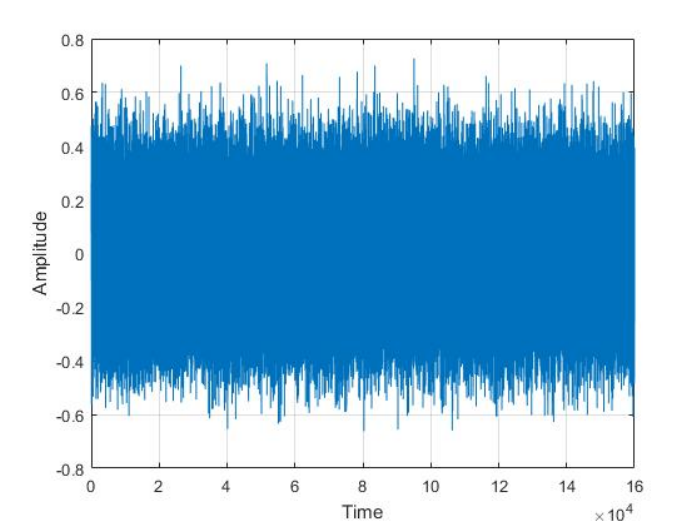} 
 \label{7a}}
   \begin{center}
      \textbf{(a)}    \end{center} 
      {\includegraphics[width=1\columnwidth]{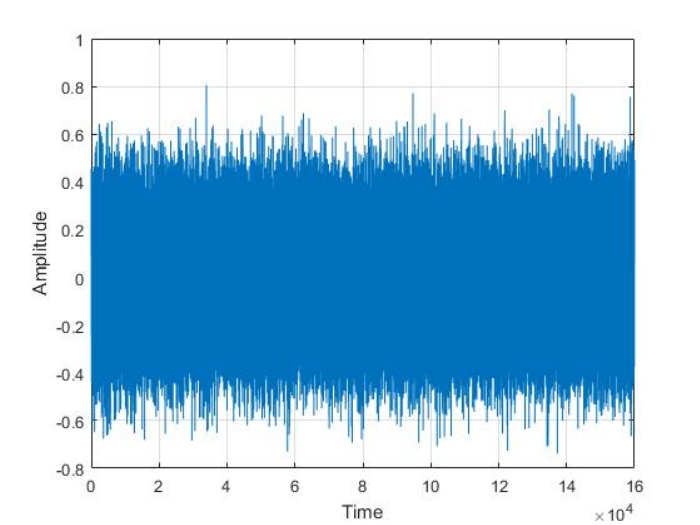}}
\label{7b}
 \begin{center}
      \textbf{(b)}   \end{center} 
    \caption{OFDM signals. (a) transmitted signal, (b) received signal.}
    \label{fg7}
\end{figure}

Fig.\ref{fg7} shows both the OFDM transmitted and received signals. Accordingly, Fig.\ref{fg7}(a) illustrates the transmitted signal, and Fig.\ref{fg7}(b) depicts the received signal after passing through a noisy channel. Note that the transmitted signal is composed of multiple subcarriers. The subcarriers are orthogonal to each other which means that they do not interfere with each other, making it possible to transmit data at a higher rate compared to other modulation techniques. However, when the signal passes through a noisy channel as shown in Fig.\ref{fg7}(b), the noise can cause errors in the received signal, leading to degradation in the signal quality which can be corrected using error correction techniques.

\subsection{HISTOGRAM ANALYSIS}
\begin{figure}[h]
{\includegraphics[width=0.4\columnwidth]{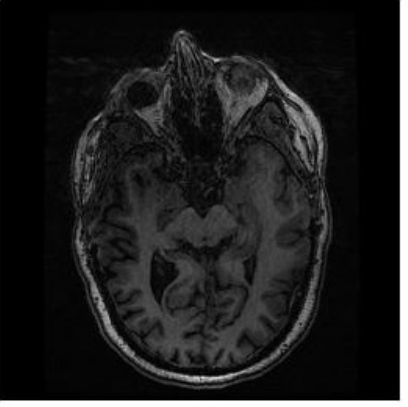} 
 \label{8a}}
      {\includegraphics[width=0.5\columnwidth]{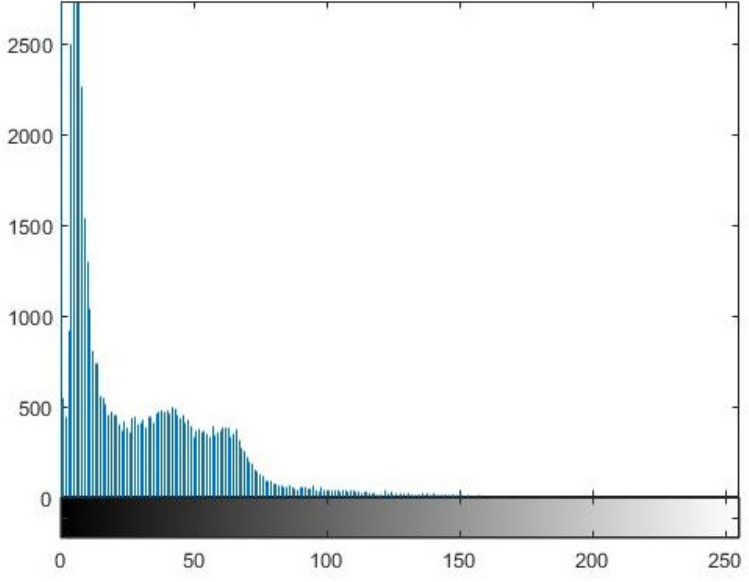}}
\label{8b}
 \begin{center}
      \textbf{(a)}   \end{center} 

      {\includegraphics[width=0.4\columnwidth]{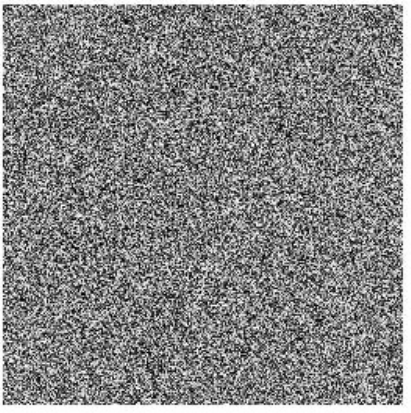} 
 \label{8a}}
      {\includegraphics[width=0.5\columnwidth]{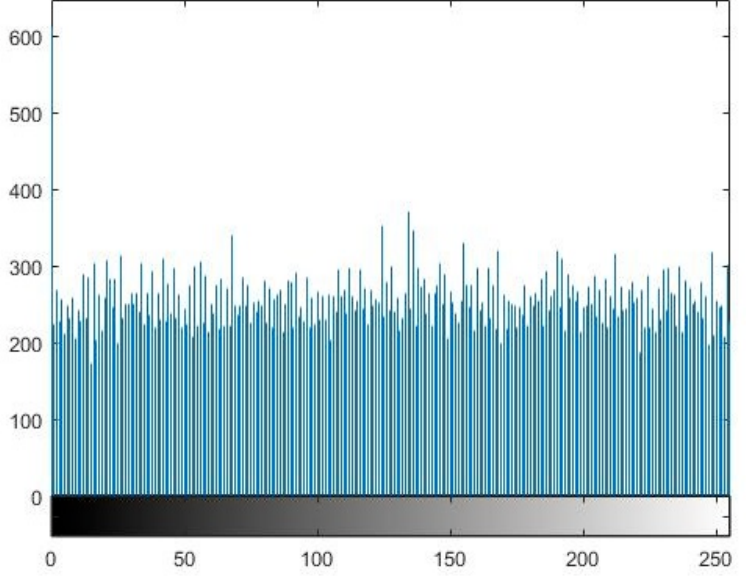}}
\label{8b}
 \begin{center}
      \textbf{(b)}   \end{center} 
      {\includegraphics[width=0.4\columnwidth]{Figures/fig8a.pdf} 
 \label{8a}}
      {\includegraphics[width=0.5\columnwidth]{Figures/fig8ahi.pdf}}
\label{8b}
 \begin{center}
      \textbf{(c)}   \end{center} 
    \caption{Histogram of: (a) original image ,(b) encrypt image and (c) decrypt image .}
    \label{fg8}
\end{figure}

A visual depiction of the distribution of image pixels is known as an image histogram, and it is utilized to evaluate image encryption algorithms. To assess the efficacy of the encryption process, it is critical to analyze the histogram of the encrypted image \cite{hosny2021new}. Fig.\ref{fg8} displays the histogram of the encrypted and decrypted images.

\subsection{NIST Statistical test }

To determine the effectiveness of an encryption algorithm, it must produce an encrypted image that is highly random. The NIST statistical test suite is a useful tool that includes statistical tests for measuring the randomness of the sequence created by the encryption algorithm. The significance level for all tests in the NIST suite is 0.01 \cite{rukhin2001statistical}. Some results of the different statistical tests are recorded in Table 2 where a P-value greater than 0.01 indicates sufficient randomness of the binary sequence. The research experiment conducted shows that the encryption algorithm has passed all tests, thus ensuring that the binary sequence produced is random.

\begin{table}[]
\centering 
\caption{NIST statistical test.}
\label{nist_table}
\fontsize{8}{10}\selectfont 
\setlength{\tabcolsep}{3pt} 
\renewcommand{\arraystretch}{1.2} 
\begin{tabular}{c c c c} 
\hline
\textbf{Index} & \textbf{Test Name}                & \textbf{P\_value} & \textbf{Result} \\ \hline
\textbf{1}     & Frequency                         & 0.220             & \textbf{Random} \\ 
\textbf{2}     & Block frequency                   & 0.468             & \textbf{Random} \\ 
\textbf{3}     & Runs                              & 0.931             & \textbf{Random} \\ 
\textbf{4}     & Longest-run-of-ones               & 0.212             & \textbf{Random} \\ 
\textbf{5}     & Discrete Fourier Transform        & 0.317             & \textbf{Random} \\ 
\textbf{6}     & Binary matrix rank                & 0.268             & \textbf{Random} \\ 
\textbf{7}     & Non overlapping template matching & 0.894             & \textbf{Random} \\ 
\textbf{8}     & Overlapping template matching     & 0.535             & \textbf{Random} \\ 
\textbf{9}     & Serial                            & 0.438             & \textbf{Random} \\ 
\textbf{10}    & Random excursion                  & 0.791             & \textbf{Random} \\ \hline
\end{tabular}    
\end{table}

\subsection{BER peformance}

\begin{figure}[]
{\includegraphics[width=0.45\columnwidth]{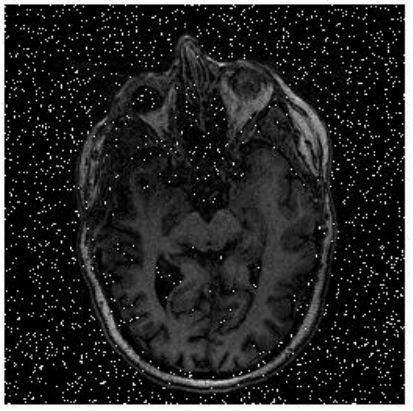} 
 \label{9a}}
      {\includegraphics[width=0.45\columnwidth]{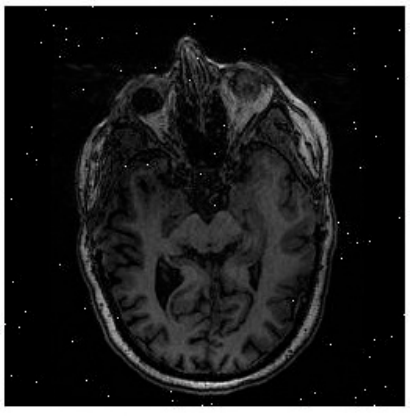}}
\label{9b}
 \begin{center}
      \textbf{(a)} \ \ \ \ \ \ \ \ \ \ \ \ \ \ \ \ \ \ \ \ \ \ \ \ \ \ \ \ \ \ \      \textbf{(b)}    \end{center} 
\begin{center}
      {\includegraphics[width=0.45\columnwidth]{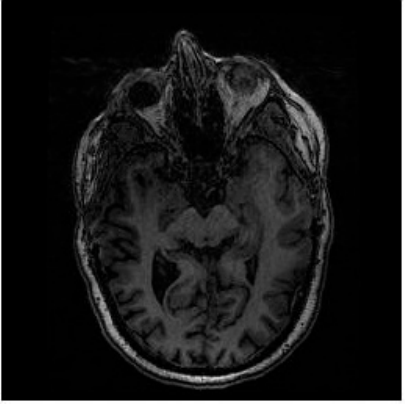}} \end{center} 
 \begin{center}
      \textbf{(c)}   \end{center} 
    \caption{Comparing reconstructed medical images at different SNR.          (a) SNR=5 db, (b) SNR=10 db and  (c ) SNR=20 db.}
    \label{fg9}
\end{figure}

Fig.\ref{fg9} shows a comparison of reconstructed medical images at different signal-to-noise ratios (SNR) of 5 dB, 10 dB, and 20 dB. The results demonstrate that as the SNR increases, the quality of the reconstructed images also improves. Additionally, Fig.\ref{fg10} illustrates the relationship between signal-to-noise ratio (SNR) and bit error rate (BER) which is a key indicator of the performance of an OFDM system in image transmission. The figure shows that as SNR increases, the BER decreases, indicating that the OFDM system performs better in terms of image transmission quality. It also highlights the importance of having a high SNR in achieving reliable image transmission using the OFDM technique. In comparison to previous work on secure image transmission, our proposed method has demonstrated superior performance in terms of bit error rate (BER). Specifically, our method achieved a lower BER of 0.005 compared to 0.007 in SNR=30db obtained in \cite{naik2017efficient}. This highlights the effectiveness of our proposed approach in improving the security and reliability of image transmission.

\begin{figure}[]
{\includegraphics[width=1\columnwidth]{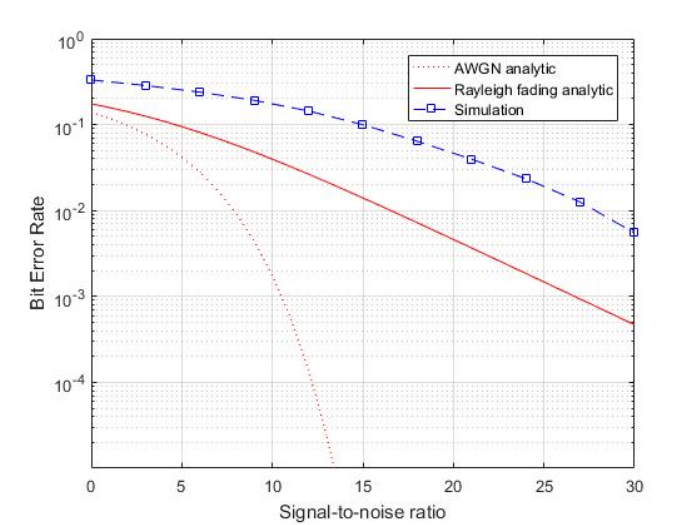} 
 \label{fg10}}
    \caption{BER performance for medical images.}
    \label{fg10}
\end{figure}

\section{Conclusion}
In conclusion, the combination of a 6-D hyperchaotic system and OFDM can be a viable option for safeguarding sensitive medical images from potential breaches and unauthorized access. In fact, the use of hyperchaotic systems can enhance security as they are very intricate and challenging to predict, making it harder for attackers to intercept or decode the transmitted data. Besides, OFDM can facilitate efficient bandwidth utilization and combat multipath fading, resulting in reliable and superior image transmission quality. Based on simulation results, this algorithm exhibits promising performance.

\bibliographystyle{IEEEtran}

\end{document}